# Graphene Molecules Contributing to the Infrared Bands of Carbon Rich Planetary Nebulae

Norio Ota[1], Aigen Li[2], Laszlo Nemes[3] and Masaaki Otsuka[4]

[1]Graduate school of Pure and Applied Sciences, University of Tsukuba, Tsukuba-City Ibaraki, 305-8571,Japan, e-mail; n-otajitaku@nifty.com,

[2]Department of Physics and Astronomy, University of Missouri, Columbia, MO 65211, USA,

[3]Research Center for Natural Sciences, Ötvös Lóránd Research Network, Budapest 1519, Hungary,

[4]Okayama Observatory, Kyoto University, Asakuchi Okayama, 719-0232, Japan,

## Abstract

It is well known since 2010 that fullerene $C_{60}$ is widespread through the interstellar space. Also, it is well known that graphene is a source material for synthesizing fullerene. Here, we simply assume the occurrence of graphene in space. Infrared spectra of graphene molecules are calculated to compare both to astronomical observational spectra and to laboratory experimental one. Model molecules for DFT calculation are selected by one astronomical assumption, that is, single void in charge neutral graphene of $C_{13}$, $C_{24}$ and $C_{54}$, resulting $C_{12}$, $C_{23}$ and $C_{53}$. They have a carbon pentagon ring within a hexagon network. Different void positions are classified as different species. Single void is surrounded by 3 radical carbons, holding 6 spins. Spin state affects molecular configuration and vibrational spectrum. It was a surprise that the triplet state is stable than the singlet. Most of charge neutral and triplet spin state species show closely resembling spectra with observed one of carbon rich planetary nebulae Tc1 and Lin49. We could assign major bands at 18.9 micrometer, and sub-bands at 6.6, 7.0, 7.6, 8.1, 8.5, 9.0 and 17.4 micrometer. It is interesting that those graphene species were also assigned in the laboratory experiments on laser-induced carbon plasma, which are analogies of carbon cluster creation in space. The conclusion is that graphene molecules could potentially contribute to the infrared emission bands of carbon-rich planetary nebulae.

## 1. Introduction

Fullerene $C_{60}$ was discovered by Kroto et al. (1985, the1996 Nobel Prize) in the sooty residues of vaporized carbon. They already suggested that "fullerene may be widely distributed in the Universe". The presence of $C_{60}$ in a wide variety of astrophysical environments was revealed by the detection of a set of emission bands at 7.0, 8.45, 17.3 and 18.9μm (Cami et al. 2010, Sellgren et al. 2010, Zhang & Kwok 2011, Bern´e et al. 2017,Garc´ıa-Hern´andez et al. 2010). Typical astronomical objects are the Galactic planetary nebula (PNe) Tc1 (Cami et al. 2010) and the Small Magellanic Cloud (SMC) PNe Lin49 (Otsuka et al. 2016), whose infrared spectra (IR) were obtained with the Infrared Spectrograph (IRS) on board the Spitzer Space Telescope. Experimentally-measured and theoretically-computed vibrational spectra of $C_{60}$ (Martin 1993, Fabian 1996, Cami et al. 2010, Candian et al. 2019) coincide with observations on Tc1 and Lin49, especially with the bands at 18.9μm and 17.4μm. However, there remain some undetermined observed bands. This motivates us to study the occurrence of other carbon species.

It is well known that graphene is a raw material for synthesizing fullerene as proposed by Kroto & McKay (1988). Chuvilin et al. (2010) showed experimentally that $C_{60}$ could be formed from a two-dimensional graphene sheet. By observation of Lin49, Otsuka et al. (2016) suggested the presence of small graphite and graphene sheets, or fullerene precursors. Graphene was first experimentally synthesized by A.K. Geim and K.S. Novoselov (2007, the 2010 Nobel Prize). The possible presence of graphene in space was reported by Garc´ıa-Hern´andez et al. (2011a, 2011b, 2012) seen in several PNe. These infrared features appear to be coincident with the transitions of planar $C_{24}$ having seven carbon hexagons. However, full observed bands cannot be explained by $C_{24}$ or hexagon network molecules. Duboscq et al (2019) tried many $C_{24}$ related molecules to study graphene group infrared spectra. Some hints come from studies on polycyclic aromatic hydrocarbon (PAH). In 2014, Ota (2014) suggested void induced PAH show calculated infrared bands closely resembling with astronomically observed unidentified infrared bands in a wide range of 3 to 20μm. He explained the origin by carbon SP3 defect among SP2 network caused by single void in carbon hexagon network. Also, Galue & Leines (2017) predicted physical model supposing $\pi$-electron irregularity among regular $\pi$-network. In this paper, we apply only one astronomical assumption, that is, single void on graphene molecule. At the first part, density functional theory (DFT) based calculation was applied to model molecules. Calculation process is, (1) to start from charge neutral molecules of small size $C_{13}$, medium $C_{24}$ and large $C_{54}$, (2) to make single carbon void resulting $C_{12}$, $C_{23}$ and $C_{53}$, (3) to distinguish void positions for every molecule as different species, (4) to calculate spin-dependent molecular configuration for singlet spin state of Sz=0/2 and triplet Sz=2/2. Thus, we will obtain infrared spectra for all neutral and spin-state dependent molecules. At the second part, we will compare such calculated spectra with astronomically observed bands of Tc1 PNe and Lin49 PNe. At the third part, we like to compare with laboratory experimented laser induced carbon plasma spectrum previously published by Nemes et al (2017). We will conclude that void induced graphene can contribute to the infrared bands of carbon rich planetary nebula.

## 2. Model Molecules

Model molecules are illustrated in Figure 1. Our selection starts from the coronene kernel-$C_{24}$ (seven carbon hexagon rings) possibly detected by Garc´ıa-Hern´andez et al. (2011a) in planetary nebulae. To find size dependence, we add a smaller one $C_{13}$ and a larger one $C_{54}$. In this paper, we apply one assumption, that is, single void on such model molecule. One capability of void creation is the cosmic ray attack on graphene, others photoionization and molecule to molecule attack. For example, as illustrated on top left of Figure 1, cosmic rays, mainly high speed proton and electron, may attack such graphene and kick out one carbon atom. Detailed origin and mechanism of void creation is not discussed in this paper. The resulting void-induced graphene molecules are $C_{12}$, $C_{23}$, and $C_{53}$, which have one carbon-pentagon-ring within the hexagon network. Molecular configuration depends on the void position as marked by a, b, c, d, e, f, in left of Figure 1. Void-induced molecule is named by suffixing a, b, and so on, such as $C_{23}$-a, $C_{23}$-b. It should be noted that initial void molecule has 3 radical carbons and holds 6 spins. Spin multiplicity should be considered to obtain precise configuration and molecular vibrational infrared spectrum. In case of single void, we should compare the singlet spin-state of Sz=0/2 and triplet one of Sz=2/2.

In the calculation, we used density functional theory (DFT) (Hohenberg& Kohn 1964, Kohn & Sham 1965) with the unrestricted B3LYP functional (Becke1993). We utilized the Gaussian09 software package (Frisch et al.2009) employing an atomic orbital 6-31G basis set (Ditchfield et al. 1971). The first step of the calculation was to obtain the self-consistent energy, optimized atomic configuration and spin density. Unrestricted DFT calculation was done to have the spin dependent configuration. The required convergence of the root-mean-square density matrix was $10^{-8}$. Based on such optimized results, harmonic vibrational frequencies and transition strengths were calculated. The vibrational intensity was obtained as the molar absorption coefficient ε(km/mol). The standard scaling is applied to the frequencies by employing a scale factor of 0.975 for pure carbon system taken from the laboratory experimental value of 0.965 on coronene molecule of $C_{24}H_{12}$ (Ota 2014). Correction due to anharmonicity was not applied to avoid uncertain fitting parameters. To each spectral line, we assigned a Gaussian profile with a full width at half maximum (FWHM) of $4cm^{-1}$.

Calculated optimized configurations are illustrated in right of Figure 1. We can notice slight configuration and energy change between molecules with spin-state of the singlet (Sz=0/2) and the triplet (Sz=2/2).

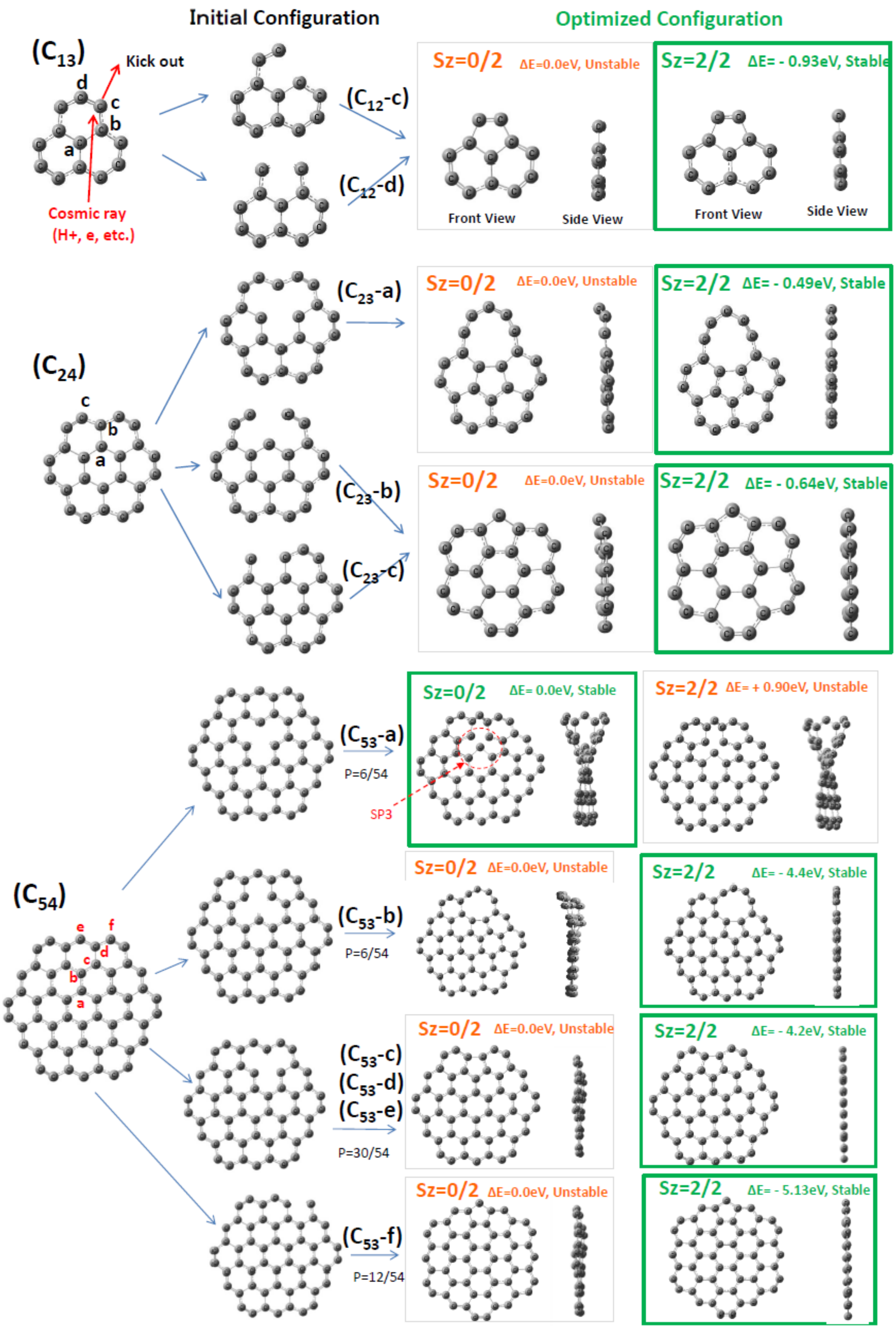

Initial Configuration
Optimized Configuration
(C13)
Kick out
Cosmic ray
(H+, e, etc.)
(C12-c)
(C12-d)
Sz=0/2 ΔE=0.0eV, Unstable
Sz=2/2 ΔE= - 0.93eV, Stable
Front View
Side View
(C24)
(C23-a)
(C23-b)
(C23-c)
Sz=0/2 ΔE=0.0eV, Unstable
Sz=2/2 ΔE= - 0.49eV, Stable
Sz=0/2 ΔE=0.0eV, Unstable
Sz=2/2 ΔE= - 0.64eV, Stable
(C54)
(C53-a)
P=6/54
Sz=0/2 ΔE= 0.0eV, Stable
SP3
Sz=2/2 ΔE= + 0.90eV, Unstable
(C53-b)
P=6/54
Sz=0/2 ΔE=0.0eV, Unstable
Sz=2/2 ΔE= - 4.4eV, Stable
(C53-c)
(C53-d)
(C53-e)
P=30/54
Sz=0/2 ΔE=0.0eV, Unstable
Sz=2/2 ΔE= - 4.2eV, Stable
(C53-f)
P=12/54
Sz=0/2 ΔE=0.0eV, Unstable
Sz=2/2 ΔE= - 5.13eV, Stable

Fig. 1.—Model graphene molecules are starting from $C_{13}$, $C_{24}$ and $C_{54}$. Assumption is single void in such neutral graphene molecule. Void position and molecule is classified by suffix a, b, and so on , as like $C_{23}$-a, $C_{23}$-b. Resulted molecule include carbon pentagon among hexagon network. It should be noted that one void bear six spins. Spin state affect molecular configuration. We should calculate multiple-spin-state even for neutral molecule. Converged molecular configuration are illustrated for both the singlet spin-state (Sz=0/2) and the triplet (Sz=2/2). Stable energy spin-state is enclosed by bold green frame.

## 3, Spin State Analysis

We tried multiple spin-state analysis for every molecule. Example is shown in Figure 2 for $C_{23}$-a and $C_{23}$-b. Initial void has three radical carbons and allows six spins as discussed in graphene-nanoribbon (Ota 2011). One radical carbon holds two spins, which are forced to be parallel up-up spins (red arrow) or down-down spins (blue) for avoiding unlimited large coulomb energy due to Hund's rule (Hund 1923). Single void creates three radical carbons, and holds six spins as illustrated in left of Figure 2. By the recombination between carbon atoms, there arises cancelling between (up-up) and (down-down) spins as imaged in middle. There remain one (up-up) spin pair to result the triplet spin-state of Sz=2/2. Detailed spin density was calculated by DFT in right at a cutting surface of $10e/nm^3$. We can see up-spin major spin cloud as explained above.

Such spin alignment relaxes atomic configuration as shown in Figure 3. We can see detailed configuration of the singlet spin state of ($C_{23}$-a, Sz=0/2). There is a slight bending of top positioned carbon. While in case of the triplet (Sz=2/2), configuration changes to a flat plane. Molecular energy was reduced by 0.49eV due to such relaxation. Similar phenomena were obtained for a case of C23-b. Other molecules' configuration and energy change are summarized in Figure 1. Energy change from the spin state singlet (Sz=0/2) to the triplet (Sz=2/2) was noted by $\Delta E$ in every column. Stable spin-state is enclosed by bold green frame. In most species, the triplet spin state is stable than the singlet. Molecular configuration was relaxed and energy was reduced. Exception is only ($C_{53}$-a) to show the stable singlet state. There arises serious configuration change as shown in Figure 1, that is, one irregular SP3-bond (marked by red circle) among SP2 network, which dissolves up-up spin pair.

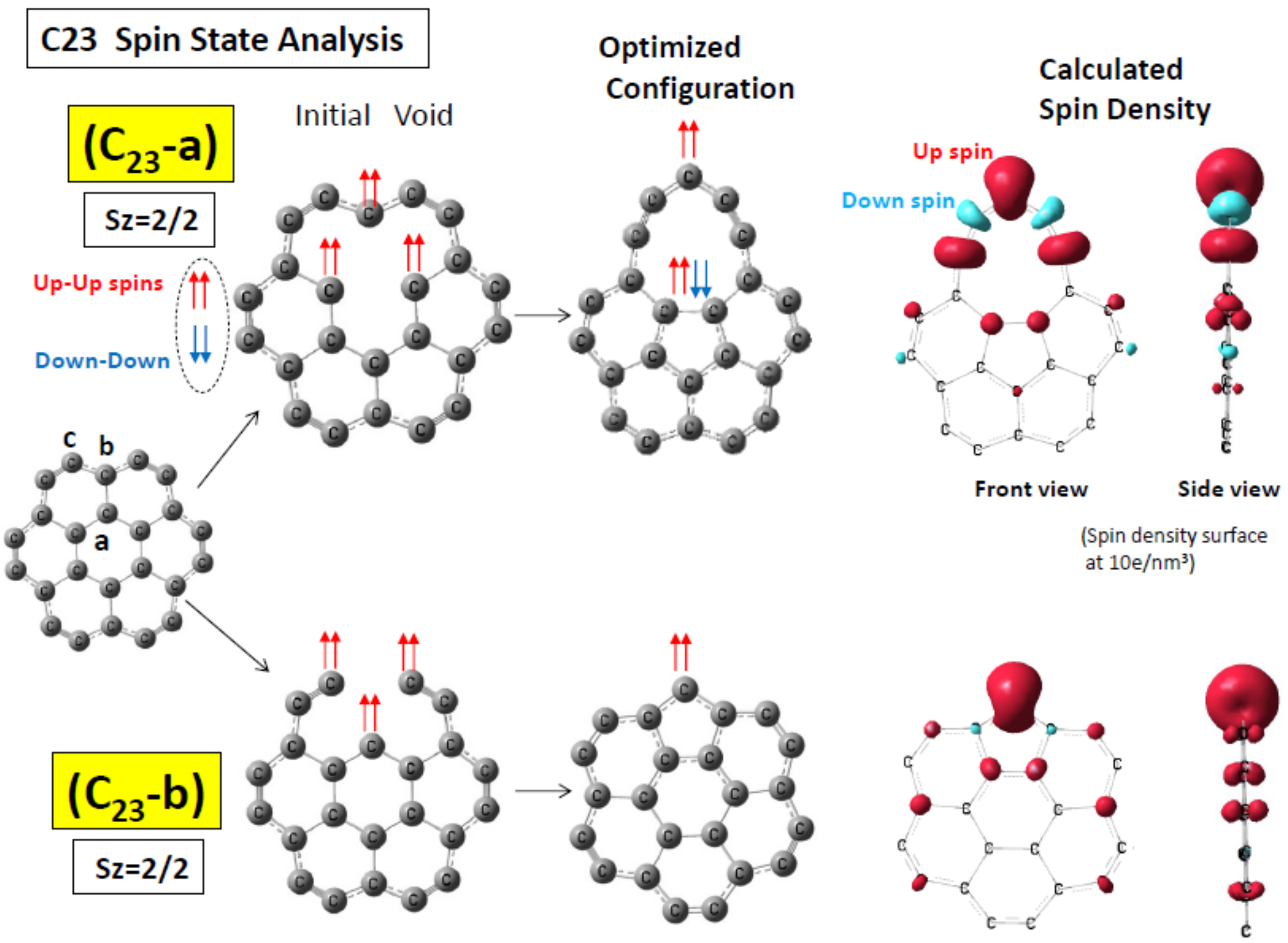

Fig. 2.— Spin state analysis for $C_{23}$-a and $C_{23}$-b. One radical carbon holds two spins, which should be parallel up-up spins (red arrows) or down-down (blue) to avoid unlimited large coulomb energy. Initial void bear three radical carbons and allow six spins. By recombination of carbon atoms, there arises cancelling of (up-up) spins by (down-down) one. There remain one (up-up) pair, resulting stable triplet spin-state (Sz=2/2). DFT calculated spin density is illustrated on right by red cloud for up-spin and blue for down-spin.

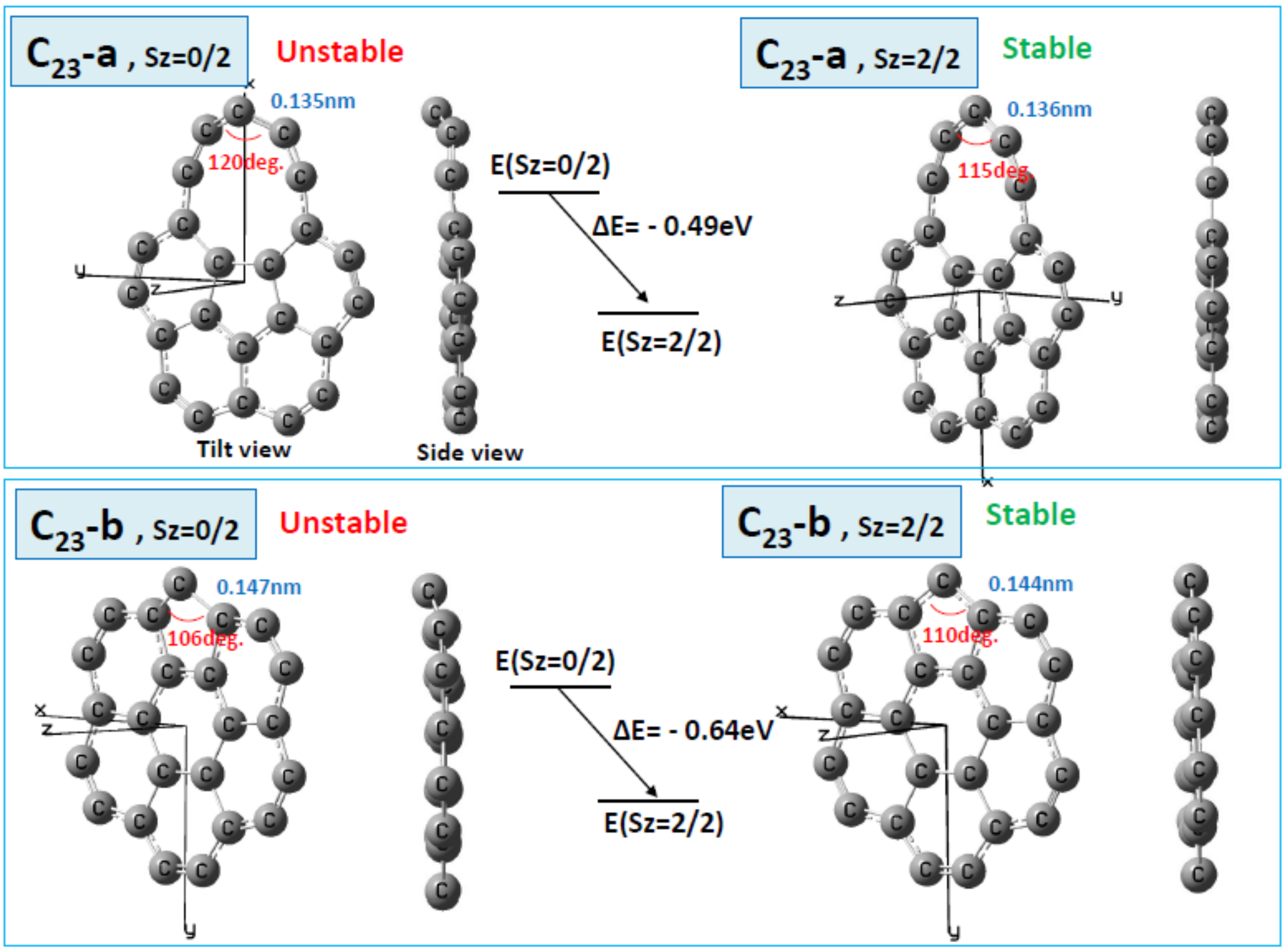

Fig. 3.—Comparison of molecular configuration and energy difference between the singlet spin-state (Sz=0/2) and the triplet (Sz=2/2) for molecules of neutral $C_{23}$-a, and neutral $C_{23}$-b. The singlet spin-state (Sz=0/2) shows bending configuration at top carbon, whereas the triplet (Sz=2/2) shows just plane configuration. Molecular energy is lower (stable) for the triplet state than the singlet by -0.49eV for $C_{23}$-a, and -0.64eV for $C_{23}$-b.

## 4, Infrared Spectrum and Fundamental Mode Analysis

Calculated infrared spectra are shown in Figure 4. Left columns are cases for the singlet spin state (Sz=0/2), while right one for the triplet (Sz=2/2). Spectrum with stable energy was enclosed by bold green frame. Astronomically observed major emission line of 18.9μm was marked by a green dotted line. The calculated spectrum of ($C_{12}$-c, Sz=2/2) [also the same configuration for $C_{12}$-d] shows a major band at 21.1μm, which is far from observed major band of 18.9μm. This is not a candidate. In cases of $C_{12}$-a and $C_{12}$-b, we cannot get calculation convergence because of complex chain-carbon like configuration. Concerning ($C_{23}$-a, Sz=2/2), we can see an irregular 9-membered ring combined with one pentagon ring and four hexagons. Calculated major band was obtained at 18.8μm close to observed one. The spectrum of ($C_{23}$-b, Sz=2/2) [the same for $C_{23}$-c] is remarkable one showing 18.9 and 19.1μm twin major peaks. In case of $C_{53}$-a, stable spin state was the singlet (Sz=0/2), of which spectrum show peaks at 21.7μm and 19.5μm, which are far from observed one. The spectrum of ($C_{53}$-b, Sz=2/2) show 19.0μm peak close to observed one. Also, ($C_{53}$-c, Sz=2/2) [the same for $C_{53}$-d, $C_{53}$-e] demonstrates major band at 18.9μm, just the same with observed one. In case of ($C_{53}$-f, Sz=2/2), we can see twin bands at 18.4 and 18.9μm. Thus, we could obtain several candidates for assigning the astronomical observations. They are all charge neutral and the triplet spin state species.

Fundamental vibrational mode of ($C_{23}$-b, Sz=2/2) was analyzed using the Gaussian09 package. There are 63 modes for 23 carbon atoms. Large intensity bands are summarized in Table 1. All modes data are listed in Appendix. Image of energy diagram is illustrated on top right of Table 1. Zero-point vibrational energy is 27634cm$^{-1}$ (=3.42eV). The lowest vibrational energy Mode-1 is 94.9cm$^{-1}$ (=108.06μm, 0.012eV). The highest Mode-63 has 2067cm$^{-1}$ (=4.96μm, 0.26eV). Vibrational behaviors of major modes are noted in right column of the table. They are carbon to carbon (C-C) in-plane (molecular plane) stretching modes. Position of stretching bonds are different for every mode, and noted by bond name of a, a', b, b' etc. as marked on a molecule structure. For example, there are two modes corresponding to observed 18.9μm band. Mode-24 is 18.8μm with strength of 78.8km/mol, which show (C-C) in-plane stretching at bonds of c, f, c', f', i, l, i', and l'. Another one, Mode-23 is 19.1μm with 92.6km/mol, (C-C) stretching at bonds of c, f, c', f', o, p, and o'. Other major modes will be discussed in the next section by comparing astronomically observed bands.

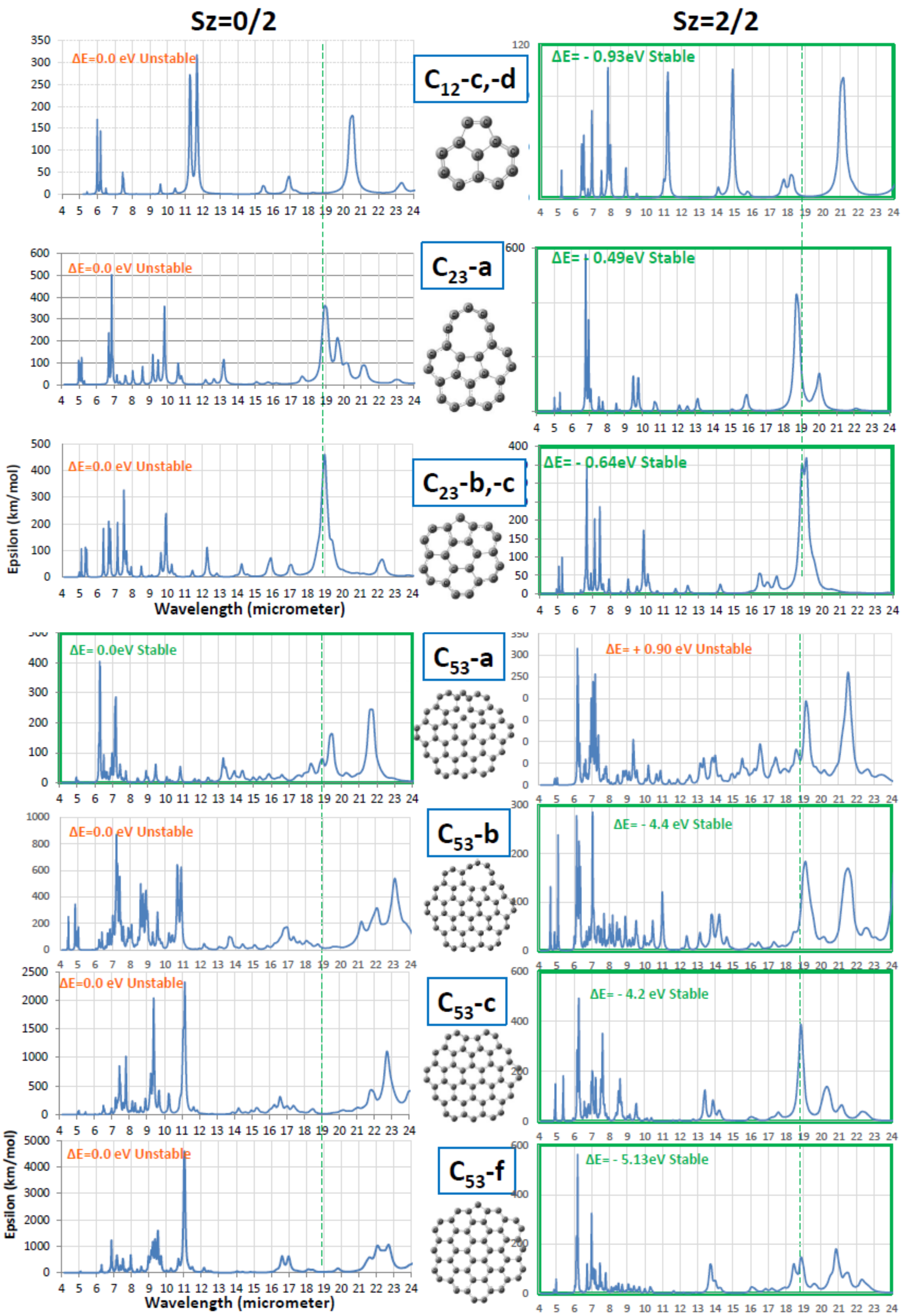
Sz=0/2
Sz=2/2
ΔE=0.0 eV Unstable
C12-c,-d
ΔE= - 0.93eV Stable
ΔE=0.0 eV Unstable
C23-a
ΔE= - 0.49eV Stable
ΔE=0.0 eV Unstable
C23-b,-c
ΔE= - 0.64eV Stable
Epsilon (km/mol)
Wavelength (micrometer)
ΔE= 0.0eV Stable
C53-a
ΔE= + 0.90 eV Unstable
ΔE=0.0 eV Unstable
C53-b
ΔE= - 4.4 eV Stable
ΔE=0.0 eV Unstable
C53-c
ΔE= - 4.2 eV Stable
ΔE=0.0 eV Unstable
C53-f
ΔE= - 5.13eV Stable
Epsilon (km/mol)
Wavelength (micrometer)

Fig. 4. — Calculated molecular vibrational infrared spectrum (absorption). Left figures are cases for the singlet spin state (Sz=0/2), while right one for the triplet (Sz=2/2). Stable spectrum is framed by bold green. Astronomically observed major emission line at 18.9μm was marked by a green dotted line.

Table-1, Major astronomically observed band, Laboratory experimental band, and DFT calculated fundamental mode of (C23-b, Sz=2/2).

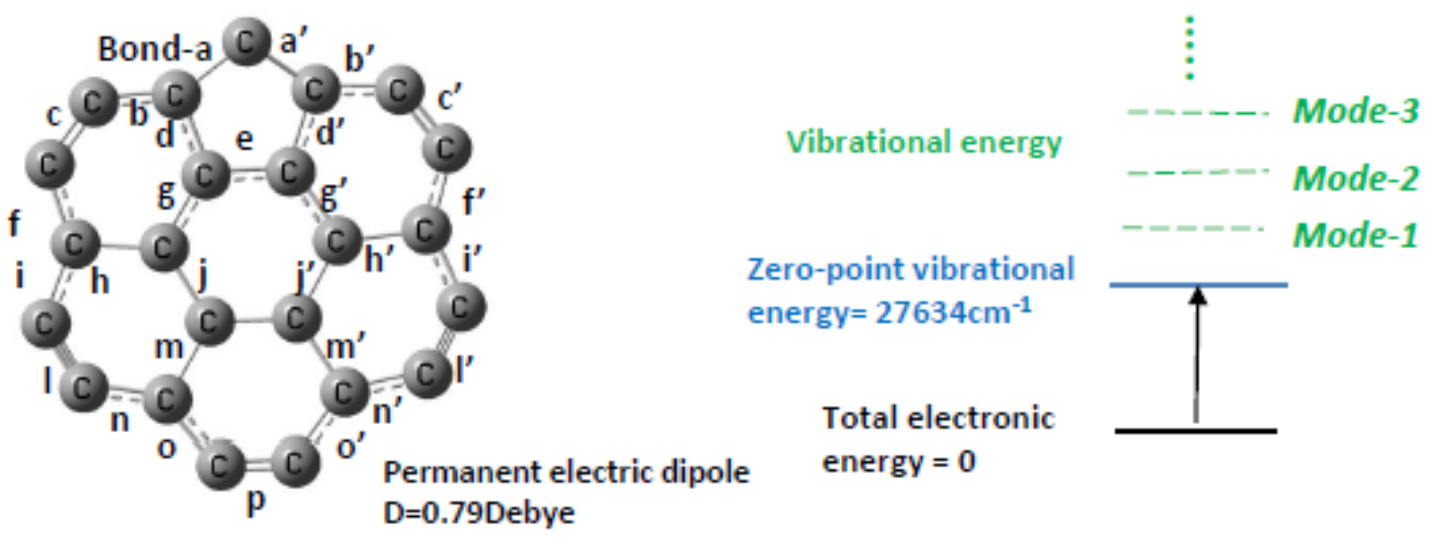

| Astronomically Observed Band (μm) | Experimental Band (laser induced carbon plasma) (μm) | Fundamental Mode of (C23-b, Sz=2/2), DFT calculation | | | |
|---|---|---|---|---|---|
| | | Mode number | Wavelength (μm) | Intensity (km/mol) | Vibrational Behavior, Carbon to carbon in-plane stretching, bond positions |
| 18.9μm | | Mode-23<br>-24 | 19.1μm<br>18.8 | 92.6<br>78.8 | c, f, c', f', i, l, i', l'<br>c, f, c', f', o, p, a o' |
| 17.4 | | -25 | 17.4 | 12.4 | c, f, c', f' |
| 16.7 | | -27 | 16.5 | 17.7 | b, b', c, c' |
| 10.0 | 10.2 | -42 | 9.9 | 52.6 | b, b', c, c' |
| 7.5 | 7.4 | -52 | 7.4 | 72.7 | a, a, a', b, b', g, g', j, j' |
| 7.0 | | -53 | 7.1 | 59.0 | a, a', b, b', d, d', g, g', j, j', o, o' |
| 6.6 | 6.7 | -55<br>-56 | 6.7<br>6.6 | 94.0<br>40.5 | a, a', b, b', f, f', i, i'<br>n, n', o, o' |

## 5, Astronomically Observed Spectra

On top of Figure 5, astronomically observed infrared spectra are illustrated for carbon rich planetary nebulae Lin49 (dark blue curve) and Tc1 (red one). We can see the largest band at 18.9μm, also a side band at 17.4μm, and shorter wavelength bands at 6.6, 7.0, 7.5, 8.1 and 8.5μm. Atomic emission lines are marked by arrows of [NeII] and [SIII].

It should be noted that the observed spectra are seen in emission. A nearby star may illuminate the molecules and excites them to give rise to infrared emission. To calculate the emission spectrum, we need information of molecular fundamental vibrational energy, intensity, excited photon energy, and thermal equilibrium conditions (Draine & Li 2001, Li & Draine 2001). We can estimate molecular information as discussed in a previous section. Problem is the lack of detailed astronomical information on Tc1 and Lin49 PNe. However, we know that, in an ideal case of ultraviolet 5eV class photoexcitation, the computed emission spectrum is almost proportional to the absorption one with opposite sign (Ota 2014, Galue et al. 2017). In this paper, calculated spectra are noted as absorbed one.

It is seen in Figure 5 that calculated absorption spectrum of the charge neutral with the triplet spin state species of (C23-b, Sz=2/2) show resemblance to the observed emission spectra of Tc1 and Lin49 PNe. As shown in Table 1, observed major band at 18.9μm could be analyzed by two modes of Mode-23 (19.1μm) and Mode-24 (18.8μm), which strength are 92.6 and 78.8km/mol. respectively. Also, observed 6.6μm band could be reproduced well by calculated two modes of Mode-55 (6.7μm) and Mode-56 (6.6μm). Also observed 7.0, 7.5, 10.0, 16.7, and 17.4μm bands could be reproduced well by Mode-53 (7.1μm), Mode-52 (7.4μm), Mode-42 (9.9μm), Mode-27 (16.5μm), and Mode-25 (17.4μm) respectively. It is amazing that such coincidence between observation and calculation was seen again in case of large molecule of ($C_{53}$-c, Sz=2/2) as shown on bottom of Figure 5. Again, we could obtain calculated modes of 7.0, 7.6, 8.5, 17.6, 18.9μm showing good coincidence. Comparing observed normalized intensity with Tc1 and Lin 49, we can see good coincidence at 18.9 and 17.4μm bands for both PNe. However, in the range of 6-9μm, band intensity of Tc1 is weaker than that of Lin49. Regarding calculated intensity of $C_{23}$-b and $C_{53}$-c, our model may support observation of Lin49. It should be noted that 6-9μm range is congeries of different carbon dust. It depends on size, charge, polycyclic, or chain. Also, there is a capability of some mixture of PAH's showing typical bands at 6.2, 7.6, 7.8, and 8.6μm. Difference between Tc1 and Lin49 comes such complex characteristics. There remain further detailed observation and enhanced model molecules.

We like to know total contribution of all model molecules, not depending on peculiar void position. Best way is the weighting sum method. In case of $C_{23}$-family, capability of void-a is 6 positions among 24 carbons of $C_{24}$ mother molecule. Weighting sum coefficient-p should be p=6/24. Also, coefficient of void-b molecule is p'=6/24 and void-c p” =12/24. Weighting sum of $C_{23}$-family should be obtained by,

$\varepsilon$ ($C_{23}$-family)= p $\varepsilon$ ($C_{23}$-a, Sz=2/2)+p’ $\varepsilon$ ($C_{23}$-b, Sz=2/2)+ p” $\varepsilon$ ($C_{23}$-c, Sz=2/2))

= (6/24) $\varepsilon$ ($C_{23}$-a, Sz=2/2)+(6/24) $\varepsilon$ ($C_{23}$-b, Sz=2/2)+ (12/24) $\varepsilon$ ($C_{23}$-c, Sz=2/2))

Weighting sum spectrum of $C_{23}$-family was illustrated in middle of Figure 6. We can see good coincidence between observation and calculation. Every weighting coefficient is noted under names of every molecule in Figure 6. Again, resulted spectrum of $C_{53}$-family could well reproduce the observed one as shown on bottom of Figure 6. It should be noted that the 6-9μm range is a complex spectral range where many modes of various carbon clusters (sizes, species, chain carbon etc.) do contribute (Duboscq 2020). Also, there is a capability of mixtures of some hydrocarbon molecules showing typical PAH bands of 6.2, 7.6, 7.8, and 8.6μm (Ota 2014). There remain further advanced studies. It is similar capability for large size $C_{53}$ –family. Especially, we can see large intensity at 6.2μm band, while cannot find on observation. Sensing range of Spitzer space telescope is 6-9μm. The 6.2μm band may be close to sensing edge with weak sensitivity. Another capability is the absorption by some PAH molecules having just 6.2μm absorption band (Ota 2014)

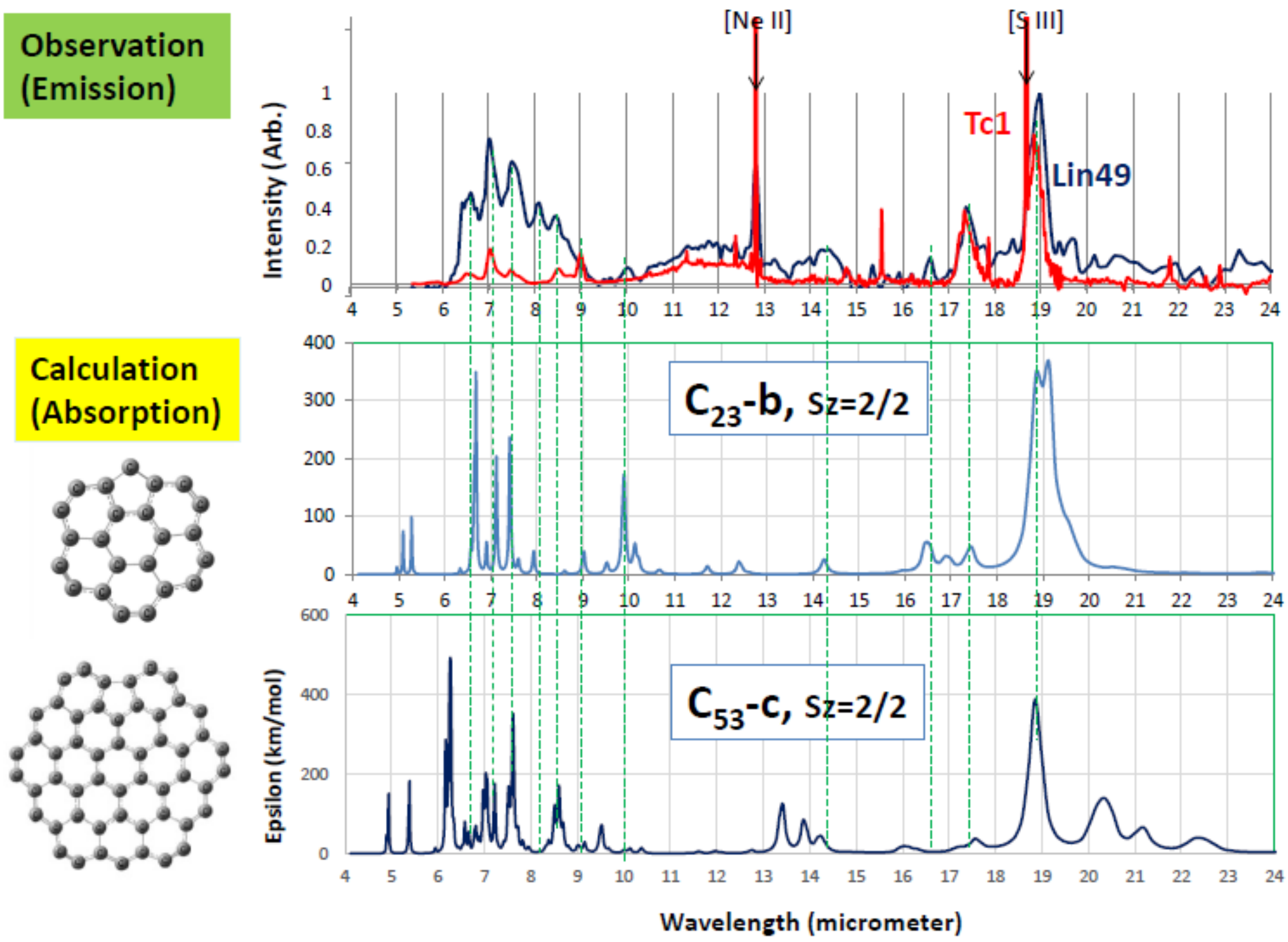

Fig. 5.— Astronomically observed emission spectrum for Tc1-PNe (on top by red curve) and Lin49 (black) compared with calculated molecular vibrational absorption spectrum of ($C_{23}$-b, Sz=2/2), and also with that of ($C_{53}$-c, Sz=2/2). Green dotted lines show major observed bands coincided well with calculated bands.

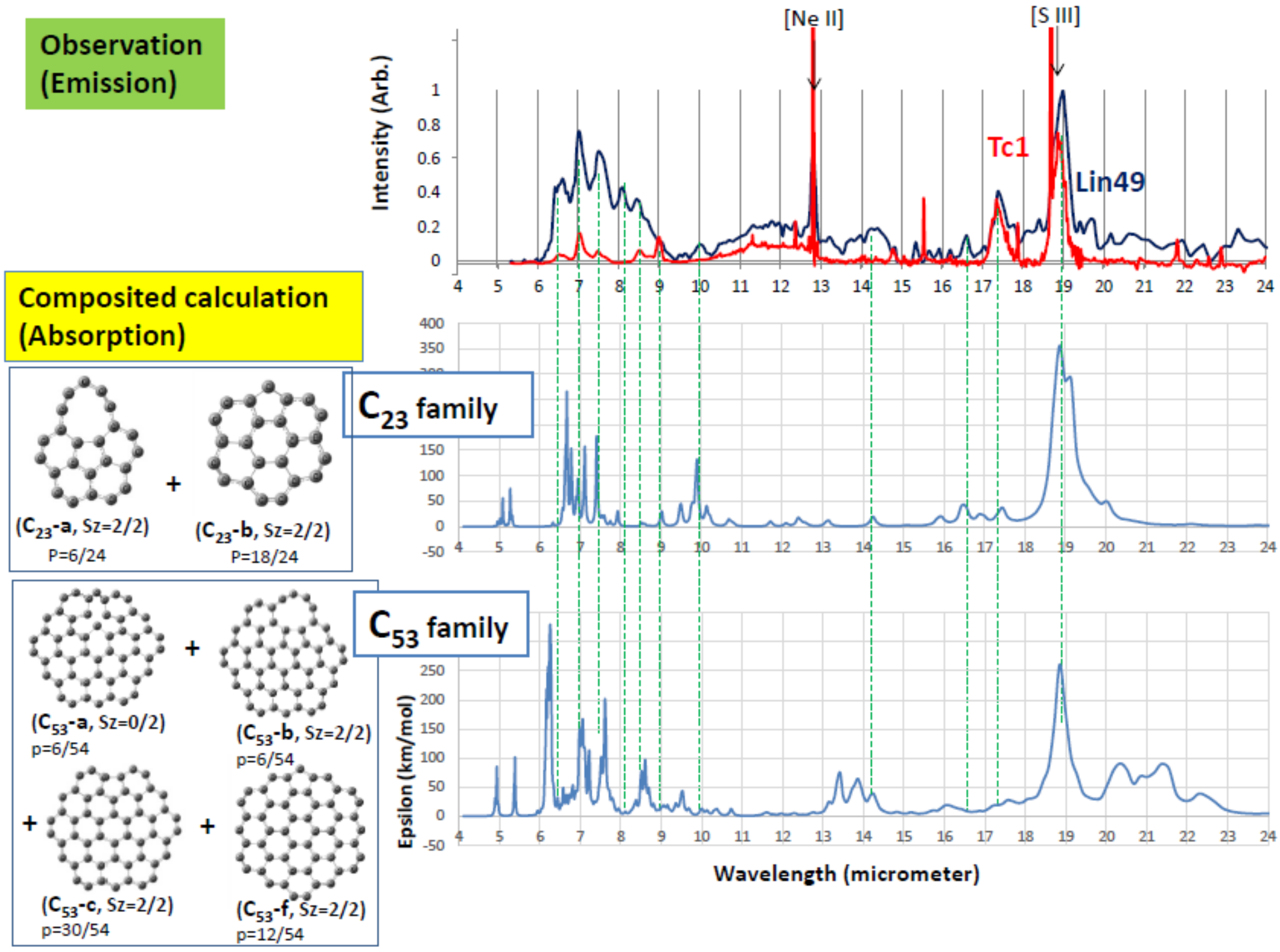

Fig. 6.—Composited calculated spectrum of $C_{23}$-family and $C_{53}$-family. Void creation capability parameter-p is considered for obtaining a weighting sum of spectra. For example, in case of $C_{23}$-a, p should be 6/24 for 6 void sites among 24 carbon of $C_{24}$. Every weighting coefficient is noted under molecule name. We can see good coincidence between observation and calculation both for $C_{23}$-family and $C_{53}$-family as shown by green dotted lines.

## 6, Laboratory Experiment of Laser Induced Carbon Plasmas

We used a laboratory experiment as an analogy of the creation of carbon dust in space, which is the laser induced carbon plasma experiment done by Nemes et al. (2017). The top panel in Figure 7 illustrates an image of the carbon dust in space. Several theories (Nozawa et al. 2003, Kozasa et al. 1987, 2009) dealt with the ejection of carbon dust from Type-IIb Supernovae. For example, in the case of a 20 times solar-mass star, after 300 days of explosion, there occur carbon dust creation in the outer helium shell during cooling of carbon plasma. Calculated average size of carbon dust is about 1nm, which is similar size with $C_{23}$ and $C_{53}$ molecules. After cooling of carbon plasma, the carbon molecules may contain graphite and graphene depending on the temperature and pressure. It is interesting that carbon dust will be further broken to smaller dust particles by backward shockwaves, that is, by sputtering with high-speed particles like proton (Nozawa et al., 2006). This may be another scenario to create voids in graphene. Carbon dust grains like graphite are also condensed in the mass-losing outflows of asymptotic giant branch (AGB) stars and planetary nebulae. Graphene could be generated from the exfoliation of graphite as a result of grain–grain collisional fragmentation (Chen et al. 2017). Also, PAHs are seen in planetary nebulae and a complete loss of their hydrogen atoms could also convert them into graphene (García-Hernández et al. 2011a, Ota 2018, Li et al 2019).

The bottom panel in Figure 7 shows the scheme of the experimental process of laser-induced carbon plasma, previously published by Nemes et al (2017). This is an analogy for carbon cluster creation in space. Bulk graphite was heated and excited by a Nd:YAG laser with wavelength of 1.064μm (1.16eV) in atmospheric pressure Ar-gas. Ablated and evaporated carbon molecules emit infrared light due to their molecular vibration. The emission was recorded by an HgCdTe detector for spectral analysis.

We have above analogy between carbon dust creation in the star explosion and the laboratory experiment. However, it should be noted that there is significant difference between them. Temperature of star explosion gas is starting from $10^5$-$10^8$ K. It will be 2000K after 300days, finally cooled as dust cloud (Nozawa et al. 2010). While, in case of laser induced plasma experiment, starting electron temperature was estimated to be 20000K, excited carbon temperature to be 4500-7000K, and finally cooled to room temperature (Nemes et al. 2005). Other detailed conditions, for example between interstellar and circumstellar dust, are out of scope in this study.

The observed infrared emission spectrum from 4 to 12μm is shown in panel (A) of Figure 8 copied from published data (Nemes et al. 2017). An intense peak is seen at 7.4μm. We compared this spectrum to the DFT calculated weighting sum spectrum of $C_{23}$-family and $C_{53}$-family in panel (B). The experimental 7.4μm peak is reproduced by the calculated one of $C_{23}$-family. We should however note another possibility for the 7.4μm line emission from neutral carbon atoms excited by population transfer from nearby excited argon atomic levels. Another feature of the observed spectrum is a plateau from 6 to 7μm, which may correspond to calculated bands at 6.7 and 7.2μm of $C_{23}$-family, and 6.3, and 7.0μm of $C_{53}$-family. Here, we remark some capabilities for 7.2-7.8μm range, that is, carbon SP3 contribution at 7.2 and 7.4μm bands as like ($C_{53}$-a), and hydrocarbon contribution as like typical PAH bands of 7.6 and 7.8μm. At a range of 5μm, we can suggest some contribution by both $C_{23}$-family and $C_{53}$-family. Also at a range of 10μm, $C_{23}$-family may contribute. Shorter wavelength bands below 5μm may come from chain carbon molecules (Nemes et al. 2005). For example, DFT calculation was tried to result the molecular vibration of chain-$C_3$ at band of 4.8μm, and chain-$C_5$ 4.6μm. A comparison to astronomical observation was shown in panel (C). Astronomically observed bands are featured by black dotted lines. We can find again good reproducibility by calculation. It should be noted that both astronomical observation and laboratory experiment could be connected very well by void induced graphene.

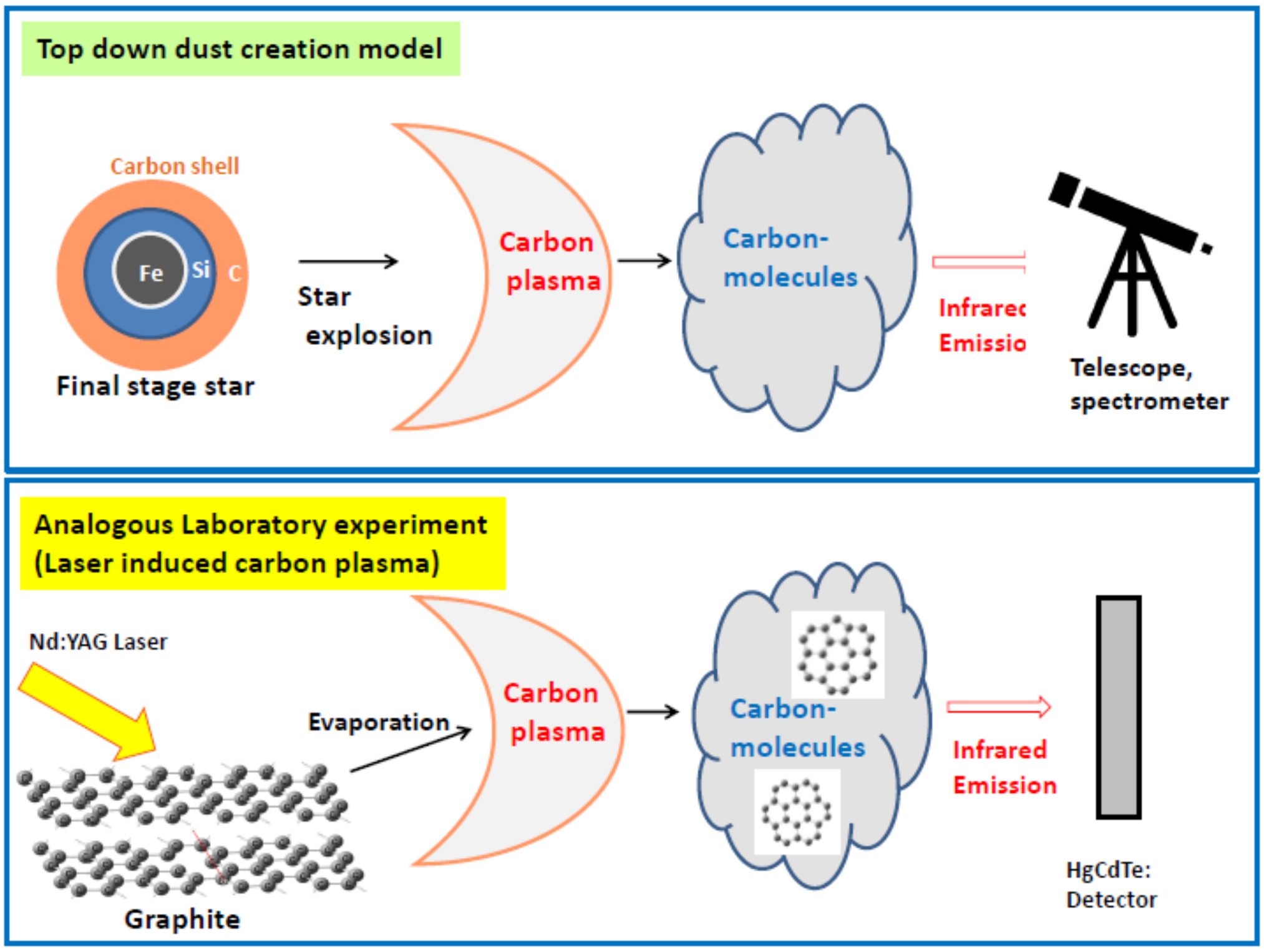

Fig. 7.—Top panel shows an assumed process of carbon dust creation in space. At stage of massive star's explosion, the outer carbon shell ejects carbon plasma to the surrounding space. Carbon plasma subsequently cools to be carbon dust. As an analogy, the laboratory experiment of laser-induced carbon plasma (Nemes et al. 2017) on bottom panel was compared. The Nd:YAG laser irradiates bulk graphite and evaporates carbon plasma resulting cooled carbon molecules. The infrared emission from carbon molecules was detected by the HgCdTe detector for spectrum analysis.

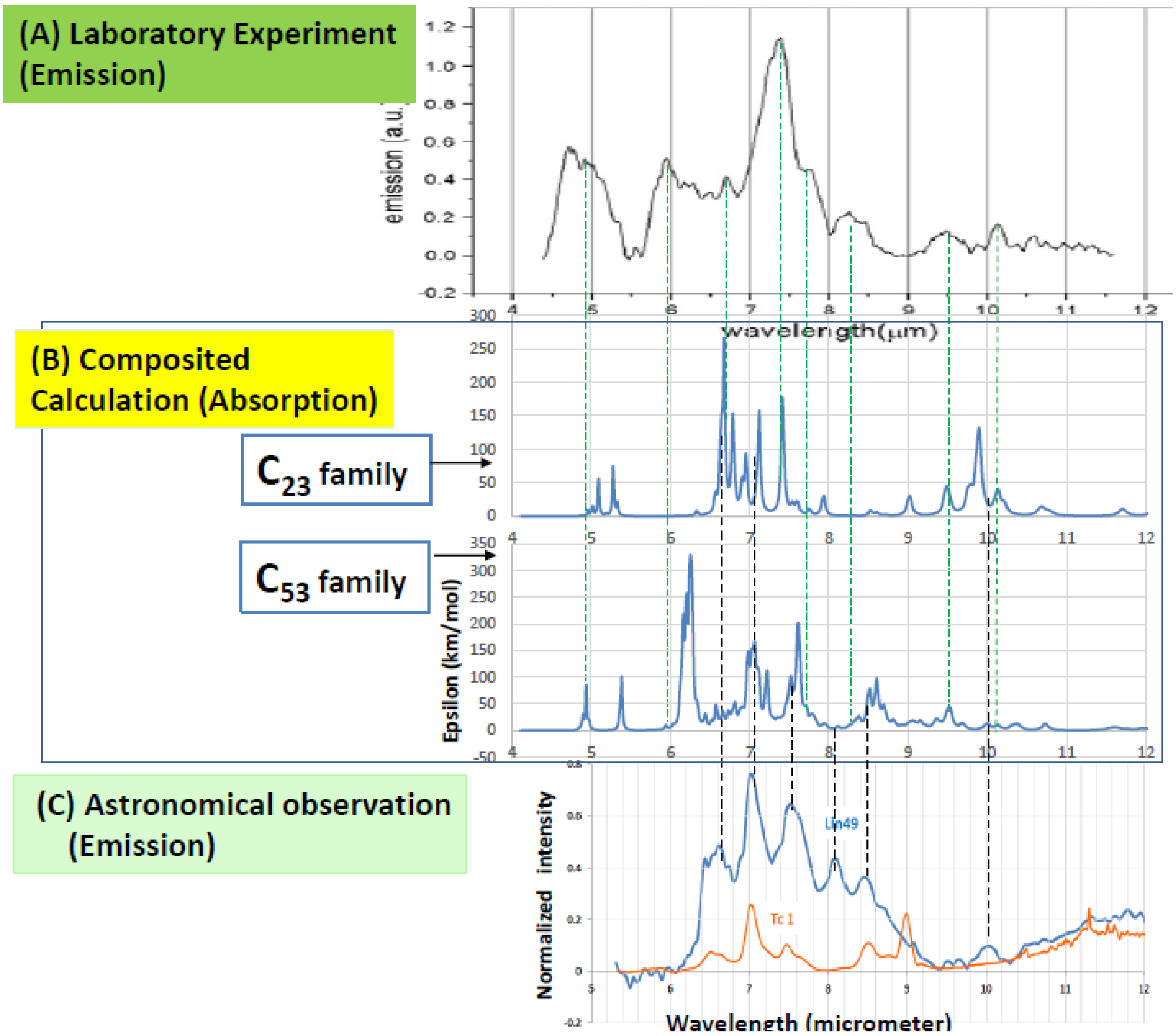

Fig. 8.—Laboratory experiment of the laser induced carbon plasma spectrum was shown in (A), which can be roughly reproduced by the weighting sum spectrum of $C_{23}$-family, also of $C_{53}$-family as illustrated in (B). The experimental bands from 6 to 7.5μm can be reproduced by the 6.3, 6.7, 7.4, 7.6μm calculated bands. We can also see band features at around 5, 8.5 and 10μm. Astronomically observed spectra are shown in (C), which is well explained by computed spectra.

## 7, Conclusion

Graphene molecules contributing to the astronomically observed infrared spectrum were investigated through DFT calculations and the laboratory experiment. Typical observations are from the Galactic Tc1 and the SMC Lin49.

(1) Starting model molecules were charge neutral $C_{13}$, $C_{24}$ and $C_{54}$ with carbon hexagon networks. Assumption of single void on such molecule is applied resulting void induced molecules of $C_{12}$, $C_{23}$, and $C_{53}$ having carbon pentagons within the hexagon network. Different void position graphene molecules are classified as different species.

(2) Single void holds three six electron spins. These spins cancel each other by recombination of carbon atoms. DFT calculation show that the triplet spin state is stable than singlet.

(3) It was found that astronomically observed emission spectra of Tc1 and Lin49 nebulae can be well explained by calculated spectra of charge neutral and the triplet spin state molecules of $C_{23}$ and $C_{53}$. We could find good coincidence for a major band at 18.9μm and six sub-bands from 6 to 8μm. Detailed fundamental vibrational mode analysis was done for $C_{23}$ with the triplet spin state.

(4) By a weighting sum method of calculated spectra for different void position molecules of $C_{23}$, we could well reproduce the observed infrared spectra. It is similar for $C_{53}$ cases. We found prominent spectral features at 6.6, 7.0, 7.5, 8.1, 8.5, 9.0, 17.4 and 18.9μm closely resembling the observed spectral details.

(5) A laboratory experiment on laser-induced carbon plasma infrared emission was examined as providing an analogy for carbon cluster creation in space. Carbon molecules arise by Nd:YAG laser ablation of graphite targets. Observed infrared emission spectra could be explained reasonably well by the presence of model molecules of $C_{23}$ and $C_{53}$.

The conclusion is that graphene molecules could potentially contribute to the infrared emission bands of carbon-rich planetary nebulae.

**Acknowledgement**

Aigen Li is supported in part by NSF AST-1311804 and NASA NNX14AF68G.

## REFERENCES

Becke, A. 1993, J. Chem. Phys., 98, 5648

Bern´e, O, &Tielens, A.G.G.M. 2012, PNAS, 109, 401

Bern´e, O., Cox, N. L. J., Mulas, G., &Joblin, C. 2017, A&A, 605, L1

Cami, J., Bernard-Salas, J., Peeters, E., &Malek, S. E. 2010, Science, 329, 1180

Candian, A., Rachid, M. G., MacIssac, H., Staroverov, V. N., Peeters, E., and Cami, J., 2019, preprint from web site of ResearchGate by Allessandra Candians, titled "Searching for stable fullerenes in space with computational chemistry"

Chen, X.H., Li, A., & Zhang, K. 2017, ApJ, 850, 104

Chuvilin, A., Kaiser, U., Bichoutskaia, E., Besley, N. A., &Khlobystov, A. N. 2010, Nature Chem., 2, 450

Ditchfield, R., Hehre, W., &Pople, J. 1971, J. Chem. Phys., 54,724

Draine, B. T. and Li Aigen 2001, ApJ 551, 807

Duboscq, C., Caldvo, F., Rapacioli, M., Dartois, E., Pino, T., Falvo, C., and Simon, A., , 2020, Astronomy & Astrophysics 634, A62

Fabian, J., 1996, Phys. Rev. B53, 13864.

Foing, B. H., & Ehrenfreund, P. 1994, Nature, 369, 296

Frisch, M. J., Trucks, G. W., Schlegel, H. B., et al. 2009, Gaussian 09, Revision B01 (GaussianInc., Wallingford, CT)

Galue, H., and Leines, G., 2017, Physical Review Lettters 119, 171102

Garc´ıa-Hern´andez, D. A., Manchado, A., Garc´ıa-Lario, P., et al. 2010, ApJL, 724, L39

Garc´ıa-Hern´andez, D. A., Kameswara Rao, N., & Lambert, D. L. 2011a, ApJ, 729, 126

Garc´ıa-Hern´andez, D. A., Iglesias-Groth, S., Acosta-Pulido, J. A., et al. 2011b, ApJL, 737, L30

Garc´ıa-Hern´andez, D. A., Villaver, E., Garc´ıa-Lario, P., et al. 2012, ApJ, 760, 107

Geim, A. K., &Novoselov, K. S. 2007, Nature Materials, 6, 183

Hohenberg, P., & Kohn, W. 1964, Phys. Rev. B, 136, 864

Hund, F., 1923, Z. Phys. 33, 345

Kohn, W., & Sham, L. 1965, Phys. Rev. A, 140, 1133

Kozasa T., Hasegawa H., 1987, Prog. Theor. Phys. 77, 1402

Kozasa T., et al., 2009, ASP Conf. Ser. 414, Cosmic Dust-Near and Far, p-43

Kroto, H. W., Heath, J. R., Obrien, S. C, Curl, R. F., and Smalley, R. E. 1985, Nature, 318, 162

Kroto, H. W., & McKay, K. 1988, Nature, 331, 328

Li Aigen and Draine B. T. 2001, ApJ 554, 778

Li, Q., Li, A., & Jiang, B.W. 2019, MNRAS, in press (arXiv:1909.12210)

Martin, M. C., Koller, D. and Mihaly, L., 1993, Phys. Rev. B47, 14607

Nemes, L., Keszler, A., Hornkohl, J., and Parigger, C., 2005, Applied Optics, 44-18, 3661

Nemes, L., Brown, E. Yang, S. C., Hommerrich, U., 2017, SpectrochimicaActa Part A Molecular and Biomolecular Spectroscopy, 170, 145

Nozawa, T., Kozasa, T., Umeda, H., Maeda, K., & Nomoto, K. 2003, *ApJ*, 598, 785

Nozawa T., Kozasa T., Habe A., 2006, ApJ 648, 435

Nozawa, T., Kozasa, T., Tominaga, N., Umeda, H., Maeda, K., Nomoto, K. and Krause, O. 2010, *ApJ*, 713, 356

Ota, N., Gorjizadeh, N., and Kawazoe, Y., 2011, Journal of Magnetics Society of Japan 35, 414, Additional data for graphene nano ribbon by Ota, N., arXiv1408.6061.

Ota, N., 2014, arXiv1412.0009. Additional data for scaling factor on arXiv1502.01766, for emission spectrum calculation on arXiv1703.05931, for SP3 defect mechanism on arXiv1808.01070.

Ota, N. 2018, arXiv, 1803.09035. Additional data on arXiv1811.05043.

Otsuka, M., Kemper, F., Leal-Ferreira, M. L., I. Aleman, M. L., Bernard-Salas, Cami, J., Ochsendorf, B., Peeters, E., and Scicluna, P. 2016, MNRAS, 462, 12

Sellgren, K., Werner, M. W., Ingalls, J. G., Smith, J. D. T., Carleton, T. M., and Christine Joblin, 2010, ApJL, 722, L54

Zhang, Y., & Kwok, S. 2011, ApJ, 730, 126

Appendix—Fundamental vibrational mode of (C23-b, Sz=2/2).

| Number of vibration mode | Calculated Vibrational Energy (cm-1) | Wavelength after scaling (μm) | Vibrational Intensity (km/mol) | Vibration behavior |
|---|---|---|---|---|
| 1 | 94.92 | 108.06 | 0.12 | |
| 2 | 115.07 | 89.13 | 0.00 | |
| 3 | 124.53 | 82.36 | 0.00 | |
| 4 | 224.61 | 45.66 | 0.01 | |
| 5 | 285.84 | 35.88 | 0.00 | |
| 6 | 327.70 | 31.30 | 0.00 | |
| 7 | 343.62 | 29.85 | 0.15 | |
| 8 | 372.84 | 27.51 | 1.34 | |
| 9 | 377.47 | 27.17 | 0.00 | |
| 10 | 381.99 | 26.85 | 0.01 | |
| 11 | 398.96 | 25.71 | 0.19 | |
| 12 | 431.32 | 23.78 | 0.71 | |
| 13 | 463.10 | 22.15 | 0.24 | |
| 14 | 476.60 | 21.52 | 0.00 | |
| 15 | 488.38 | 21.00 | 0.12 | |
| 16 | 489.53 | 20.95 | 0.00 | |
| 17 | 493.52 | 20.78 | 0.83 | |
| 18 | 498.94 | 20.56 | 0.30 | |
| 19 | 499.09 | 20.55 | 0.99 | |
| 20 | 507.37 | 20.21 | 0.00 | |
| 21 | 522.30 | 19.64 | 7.76 | |
| 22 | 526.72 | 19.47 | 11.39 | C-C in-plane in-phase shrinking |
| 23 | 536.85 | 19.10 | 92.59 | C-C in-plane stretching (c, f, c', f', o, p, o') |
| 24 | 544.47 | 18.84 | 78.84 | C-C in-pl. Str. (c, f, c', f', i, l, i', l') |
| 25 | 588.80 | 17.42 | 12.40 | C-C in-pl. twisting (c, f, c', f') |
| 26 | 606.27 | 16.92 | 7.84 | |
| 27 | 622.21 | 16.48 | 17.66 | C-C in-pl. twisting (b, b', c, c') |
| 28 | 636.09 | 16.12 | 0.00 | |
| 29 | 642.61 | 15.96 | 1.00 | |
| 30 | 650.38 | 15.77 | 0.07 | |
| 31 | 701.82 | 14.61 | 0.00 | |
| 32 | 709.64 | 14.45 | 0.15 | |
| 33 | 720.14 | 14.24 | 7.38 | |
| 34 | 726.72 | 14.11 | 0.00 | |
| 35 | 813.35 | 12.61 | 0.66 | |
| 36 | 826.87 | 12.40 | 6.85 | |
| 37 | 875.97 | 11.71 | 4.04 | |
| 38 | 938.27 | 10.93 | 0.08 | |
| 39 | 961.77 | 10.66 | 2.31 | |
| 40 | 1004.43 | 10.21 | 5.31 | |
| 41 | 1012.67 | 10.13 | 13.43 | |
| 42 | 1037.08 | 9.89 | 52.56 | C-C in-pl. stretching (f, i, h, f',i', h') |
| 43 | 1077.49 | 9.52 | 6.01 | |
| 44 | 1136.31 | 9.03 | 0.15 | |
| 45 | 1137.50 | 9.02 | 12.63 | C-C in-pl. str. (a, a', h, h',m, m') |
| 46 | 1191.71 | 8.61 | 1.80 | |
| 47 | 1233.99 | 8.31 | 0.21 | |
| 48 | 1293.21 | 7.93 | 12.36 | C-C in-pl. str. (k, n, n',o,o') |
| 49 | 1319.27 | 7.77 | 0.02 | |
| 50 | 1340.74 | 7.65 | 0.41 | |
| 51 | 1349.75 | 7.60 | 7.75 | |
| 52 | 1382.86 | 7.42 | 72.71 | C-C in-pl. str. (a, a', b, b', g, g', j, j' ) |
| 53 | 1440.52 | 7.12 | 59.05 | C-C in-pl. str. (a, a', b, b', d, d', g, g', j, j' , o, o' ) |
| 54 | 1485.22 | 6.91 | 16.04 | |
| 55 | 1536.09 | 6.68 | 93.96 | C-C in-pl. str.(a, a', b, b', f, f', i, i' ) |
| 56 | 1545.27 | 6.64 | 40.48 | C-C in-pl. str. (n, n', o, o' ) |
| 57 | 1561.68 | 6.57 | 9.82 | |
| 58 | 1619.17 | 6.33 | 2.74 | |
| 59 | 1921.21 | 5.34 | 0.48 | |
| 60 | 1942.84 | 5.28 | 30.59 | C-C in-pl. str. (c, c', f, f' , l, l', p ) |
| 61 | 2013.76 | 5.09 | 25.32 | C-C in-pl. str.(c, c', f, f' , l, l', n, n',p ) |
| 62 | 2050.50 | 5.00 | 1.31 | |
| 63 | 2067.79 | 4.96 | 3.47 | |